\documentclass[preprint2]{aastex}

\newcommand{\fkaqr}{FK\,Aqr}
\newcommand{\wolf}{V1054\,Oph}
\newcommand{\adleo}{AD\,Leo}
\newcommand{\tdel}{{\delta}t}

\slugcomment{accepted for publication in ApJ}

\shorttitle{Flares on Low-mass Stars}
\shortauthors{Kashyap et al.}

\received{2002 March 25}
\begin{document}

\title{Flare Heating in Stellar Coronae}

\author{Vinay L.\ Kashyap and Jeremy J.\ Drake}
\affil{Harvard-Smithsonian Center for Astrophysics, 60 Garden Street, Cambridge, MA 02138}
\email{vkashyap@cfa.harvard.edu, jdrake@cfa.harvard.edu}

\and

\author{Manuel G\"{u}del and Marc Audard}
\affil{Laboratory for Astrophysics, Paul Scherrer Institute, W\"{u}renlingen and Villigen, 5232 Villigen PSI, Switzerland}
\email{guedel@astro.phys.ethz.ch, audard@astro.phys.ethz.ch}

\begin{abstract}

An open question in the field of solar and stellar astrophysics
is the source of heating that causes stellar coronae to reach
temperatures of millions of degrees.  One possibility is that the
coronae are heated by a large number of small flares.  On the Sun,
flares with energies as low as those of microflares are
distributed with energy as a power\,law
of the form $\frac{dN}{dE} \propto E^{-\alpha}$ with
$\alpha\approx1.8$, and $\alpha$ appears to increase to values
2.2-2.7 for flares of lower energy.  If the slope exceeds the
critical value of 2, then in principle the entire coronal energy
input may be ascribed to flares that are increasingly less energetic,
but are more numerous.  Previous analyses of flares in light-curves of
active stars have shown that this index is generally $>2$,
%(e.g., Audard et al.\ 2000),
though it may be as low as 1.6 when strong flares alone are considered.

Here we investigate the contribution
of very weak flares, covering the milliflare energy range,
to the coronal luminosity of low-mass active stars.
We analyze {\sl EUVE}/DS events data from \fkaqr, \wolf, and \adleo\
and conclude that in all these cases the coronal emission is dominated
by flares to such an extent that in some cases the entire emission
may be ascribed to flare heating.
We have developed a new method to directly model for the first time
stochastically produced flare emission, including undetectable flares,
and their effects on the observed photon arrival times.
We find that
$\alpha_{\rm\fkaqr} = 2.60 \pm 0.34$,
$\alpha_{\rm\wolf} = 2.74 \pm 0.35$,
$\alpha_{\rm\adleo} = 2.03 - 2.32$,
and that the flare component accounts for a large fraction (generally
$>50\%$) of the total flux.

\end{abstract}


\section{Introduction}
\label{s:intro}

The source of heating of solar and stellar coronae still eludes
understanding even after decades of study (e.g., Schrijver et al.\ 1999).
Despite significant evidence that magnetic activity is the prime
driver for transferring energy into the corona, the mechanism by
which this transfer occurs is not established in either the case
of the Sun or other stars (see e.g., Rosner, Golub, \& Vaiana 1985,
Narain \& Ulmschneider 1996).
Numerous heating mechanisms, such as acoustic wave dissipation
(Stepien \& Ulmschneider 1989),
%(Haisch \& Schmitt 1996)
Alfv\'en wave dissipation (Cheng et al.\ 1979, Narain \& Ulmschneider 1990),
magnetic reconnection phenomena (Parker 1988, Lu \& Hamilton 1991) have
been proposed, all of which might play some role in the overall
heating.

Recent work in the solar case has lent strong credence to the
possibility of coronal heating being dominated by small-scale
explosive events suggestive of Parker's {\sl nanoflare} model,
which is
based on magnetic reconnections releasing energies $\sim 10^{24}$
erg event$^{-1}$.  It has been well known that solar microflares
and milliflares\footnote{
Because of the vast range of flare energies encountered, the
energy ranges of the different flare types are not well defined.
We adopt the convention (see e.g., Aschwanden et al.\ 2000) that
milliflares cover the range $E \sim 10^{29-32}$ ergs, microflares
$E \sim 10^{26-29}$ ergs, and nanoflares $E \sim 10^{23-26}$ ergs.
We consider all events down to the microflare regime to be `normal'
X-ray flare events, with similar origin, parameters, and effects,
except for the differences in energy deposition.
}
are distributed in number as
power\,laws of their energy output (Lin et al.\ 1984, Hudson 1991),
\begin{equation}
\label{e:powerlaw}
\frac{dN}{dE} = k E^{-\alpha}
\end{equation}
where $E$ is the energy of the flare and $k$ is a constant.  This
relation has been verified and extended to lower energies by various
authors, but despite the near universal acceptance of the form of
the function in Equation~\ref{e:powerlaw} (e.g., Crosby, Aschwanden,
\& Dennis 1993; see Kopp \& Poletto 1993, Shimizu \& Tsuneta 1997 for
a different perspective), neither the index $\alpha$ nor the
normalization $k$ are well determined.
For instance, $\alpha = 1.6-1.8$ in the HXR to microflare energy range,
and is variously measured to lie in the range 1.8-2.9 at lower
energies
(Shimizu 1995 [$\alpha=1.5-1.6$],
Porter, Fontenla, \& Simnett 1995 [$\alpha=2.3$],
Krucker \& Benz 1998 [$\alpha=2.3-2.6$],
Parnell \& Jupp 2000 [$\alpha=2.0-2.6$],
Aschwanden et al.\ 2000 [$\alpha=1.8$],
Winebarger et al.\ 2001 [$\alpha=2.9 \pm 0.1$],
Veronig et al.\ 2002 [$\alpha=2.03 \pm 0.09$]).
Recently Aschwanden \& Parnell (2002) have used scaling laws based
on energy balance arguments to conclude that $\alpha$ must be $\sim 1.6$
on the Sun.
The precise value of $\alpha$ is of considerable interest because if
the power\,law is steep enough ($\alpha>2$), then in principle a
multitude of small impulsive events would be sufficient to account for
the energy output of the entire corona.

Here we reconsider in particular an outstanding question in stellar
X-ray astronomy, which is the nature of the apparently quiescent emission
from active stars: does this emission actually arise from a superposition
of a multitude of impulsive events (such as milliflares and microflares),
or from truly quiescent plasma?  Previous work based on detecting
flares in EUV data (see e.g., Audard et al.\ 2000) suggests that flare
contribution is indeed an important factor.  Further,
correlations of quiescent  X-ray
flux with time-averaged U-band flare flux (Skumanich 1985, Doyle \& Butler 1985)
and the synchrotron radio luminosity (G\"{u}del \& Benz 1993),
together with the similar correlations found in the solar case
(Benz \& G\"{u}del 1994) strongly suggest a link between the apparently
quiescent emission and flares.  In addition, spectroscopic evidence for
high temperature plasma ($T \gtrsim 10^7$ K) during the quiescent
phase (Butler et al.\ 1986, Kashyap et al.\ 1994, Drake 1996, G\"{u}del et
al.\ 1997, G\"{u}del 1997) indicates
that this quiescent emission could in fact be very similar to flare
emission in origin.  Thus, apparently quiescent coronae of active stars
could be composed of a continuum of small unresolved flares, presumably
distributed as power\,laws analogous to the Sun.  This view is also
supported by the double-peaked Differential Emission Measures (DEMs)
that result when an ensemble of flaring, hydrodynamically evolving
loops are modeled on active solar analogs (G\"{u}del et al.\ 1997,
G\"{u}del 1997).

The possibility of stellar coronal heating due to small flares was
considered by Ambruster, Sciortino, \& Golub (1987) who searched
for variability in {\sl Einstein} data of active stars and discussed
the contribution of low-level flaring to heating stellar coronae.
They concluded that while flaring must contribute at some level, the
evidence does not justify extending the solar power-law distributions
to the stellar microflare case.  Later studies of ensembles of strong
stellar flares seen with {\sl EXOSAT} and {\sl EUVE} have shown these
are distributed as power\,laws
with index $\alpha=1.6-1.8$ (Collura et al.\ 1988, Pallavicini
et al.\ 1990, Osten \& Brown 1999), thus ruling out low-intensity
flares as a significant contributor to the heating budget.
In contrast, using a more
sensitive method to detect fainter flares (see Crawford et al.\ 1970),
Robinson et al.\ (1995, 1999, 2001) find that for stellar chromospheric
and transition region events observed with the high-speed photometer and
the imaging spectrograph on the {\sl HST},
$\alpha \sim 1.76-2.17$ in the chromosphere of the active dMe star CN\,Leo;
$\alpha=2.25 \pm 0.1$ in the chromosphere of the dMe star YZ\,CMi;
and $\alpha \sim 2.2-2.8$ in the transition region of the dM0e flare
star AU\,Mic.
(Note however that chromospheric and transition-region flare
distributions have no known direct correspondence with the coronal
case.)
Applying a similar method to {\sl EUVE}/DS
data, and also correcting for overlaps in flares, Audard et al.\ (1999)
find that for solar analogs EK\,Dra and 47\,Cas, $\alpha\approx~2.2\pm0.2$.
This analysis was further extended by Audard et al.\ (2000) to a
larger sample of cool stars, and they find that $\alpha$ ranging from
1.5 to 2.6, with the majority of the measurements having $\alpha>2$.
Similar results are obtained for AD\,Leo (G\"{u}del et al.\ 2001,2002).

Note that the above studies are limited to relatively large
flares ($E \gtrsim 10^{31}$ ergs) because of instrument
sensitivity, and also because the more numerous weaker flares
are harder to detect in the presence of ``contamination''
by other weak flares.  Thus, the low-energy end of the
flare distribution is subject to large uncertainties.

We have developed a new method to {\sl model} the undetectable
stellar flares and thus derive estimates of flare indices covering
the milliflare regime as well as directly estimating
the flare contribution to the observed flux.  We apply this
method to active low-mass stars \fkaqr, \wolf, and \adleo.
The datasets used are described in \S\ref{s:data}.
The analysis method is detailed in \S\ref{s:analyz}
(a glossary of the terms used is given in Appendix~\ref{s:glossary}).
The results are given in \S\ref{s:result}, and
are summarized in \S\ref{s:summary}.

\section{Data}
\label{s:data}

\begin{figure}[htb!]
%\plotone{FKAqrlc.eps}
\plotone{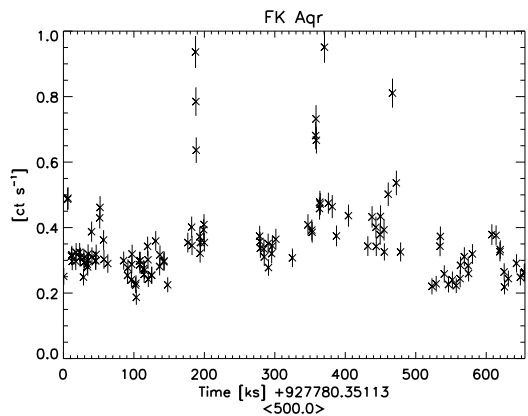}
\caption{{\sl EUVE}/DS light curve of \fkaqr.  The light-curve corrected for
instrumental effects is shown at a bin size of 500 s.  The vertical
bars denote the 1-sigma error on the count rate.
%{***The average count rate (solid line) and $1\sigma$ errors on it
%(dotted lines) are also shown.***}
Note that there are many obvious flares visible in the light
curve, in addition to a base emission which also is highly variable.}
\label{f:fkaqr_lc}
\end{figure}

Here we analyze data from the {\sl Extreme Ultra-Violet Explorer}
satellite's {\sl Deep Survey} photometer ({\sl EUVE}/DS) of 3 active
low-mass stars.\footnote{
The intrinsic {\sl EUVE} time resolution is 8 ms, and this is adequate
to resolve the sources even at the maximum count rate seen in our
observations (3 ct s$^{-1}$).
The {\sl EUVE}/DS covers a useful spectral range of 52-246 \AA, with
a peak effective area of 28 cm$^{-2}$ at 91 \AA.  This wavelength range
includes many lines from highly ionized Fe\,XVIII to Fe\,XXIII normally
found in the coronae of active stars, in addition to bound-free and
free-free continua: plasma temperatures from $\approx 1$ to 30 MK are
thus accessible for observation.
}
These stars are known to have significant flare activity, and do not
undergo eclipses, and so are amenable to straightforward modeling:

\paragraph{\fkaqr} is a BY\,Dra type, spectroscopic, double-lined,
non-eclipsing, low-mass, active binary (Table~\ref{t:fkaqr}).  Its
flare energy output has remained steady over long intervals (8 years;
Byrne et al.\ 1990), and optical modulation due to spot activity has
been observed.
Thus it is possible that flare heating could be a significant component
of coronal emission for this star.
Indeed, the {\sl EUVE}/DS light curve shows evidence of a
number of flare events (Figure~\ref{f:fkaqr_lc}) as well as large
stochastic variability in the apparently quiescent emission.

%\clearpage
\begin{table}[htb!]
\begin{center}
\caption{{\bf \fkaqr} % Low-mass, non-eclipsing, spectroscopic binary
\label{t:fkaqr}}
\begin{tabular}{l l}
\tableline
\tableline
Other Names & Gl\,867A / HD\,214479 \\
(RA, Dec)$_{2000}$ & (22:38:45.56, -20:37:16.1) \\
Spectral Type~$^a$ & dM2e/dM3e \\
Period~$^a$ & 4.08 days \\
Distance~$^b$ & 8.64 pc \\
$m_V$~$^b$ & $9^m.06$ \\
$B-V$~$^b$ & 1.47 \\
$L_X$~$^c$ & $1.3 \times 10^{29}$ ergs s$^{-1}$ \\
{\sl EUVE}/DS & 1997-oct-17 to 1997-oct-24 \\
\hfil & (130.7 ks) \\
%{***Count rate & $0.35 \pm 0.03$ ct s$^{-1}$ \\
%Background & $\sim 0.02$ ct s$^{-1}$ \\
%following numbers are Primbsch corrected and direct from event list***}
Count rate & $0.36 \pm 0.033$ ct s$^{-1}$ \\
Background & $\sim 0.023$ ct s$^{-1}$ \\
\tableline
\end{tabular}
\tablenotetext{a}{as in Strassmeier et al.\ (1993)}
\tablenotetext{b}{from the Hipparcos catalog (Perryman et al.\ 1997)}
\tablenotetext{c}{in the {\sl Einstein} energy band (0.1-4.5 keV) (Dempsey et al.\ 1993)}
\end{center}
\end{table}

%\clearpage

%\clearpage

\begin{figure}[htb!]
%\plotone{v1054Ophlc.eps}
\plotone{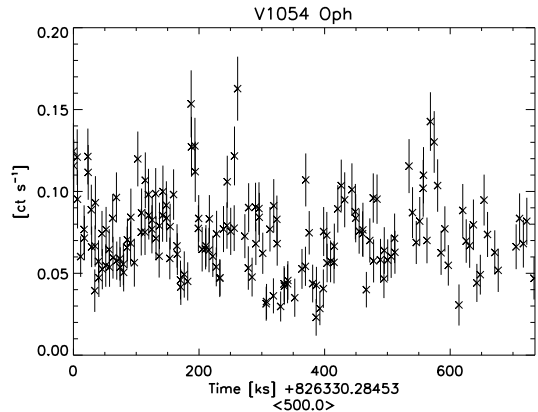}
\caption{Same as Figure~\ref{f:fkaqr_lc}, but for \wolf.
In this light curve it is difficult to cleanly distinguish a
flare event from the underlying continuous emission.}
\label{f:gj644_lc}
\end{figure}

\paragraph{\wolf} is a low-mass astrometric binary (Table~\ref{t:gj644}).
Previous X-ray observations with {\sl ROSAT} have confirmed the existence
of almost continuous flaring (Kellet \& Tsikoudi 1999) and the existence
of high-temperature plasma that is responsible for most of the intensity
variations (Giampapa
et al.\ 1996).  {\sl EUVE} data show a large number of relatively weak flare
events that blend into the variations in the quiescent emission
(Figure~\ref{f:gj644_lc}).  Such a dataset is very difficult to analyze
by the traditional means of detecting and counting flares, but poses
no difficulty to a direct modeling approach as is described below.

%\clearpage
\begin{table}[htb!]
\begin{center}
\caption{{\bf \wolf} % Low-mass flare star
\label{t:gj644}}
\begin{tabular}{l l}
\tableline
\tableline
Other Names & Wolf 630 / Gl\,644 / HD\,152751 \\
(RA, Dec)$_{2000}$ & (16:55:28.76, -08:20:10.8) \\
Spectral Type~$^a$ & M3Ve \\
Distance~$^a$ & 5.73 pc \\
$m_V$~$^a$ & $9^m.02$ \\
$B-V$~$^a$ & 1.553 \\
$L_X$~$^b$ & $5.6 \times 10^{28}$ ergs s$^{-1}$ \\
{\sl EUVE}/DS & 1994-jul-30 to 1994-aug-08 \\
\hfil & (127.9 ks) \\
%{***Count rate & $0.074 \pm 0.014$ ct s$^{-1}$ \\
%Background & $\sim 0.013$ ct s$^{-1}$ \\
%following numbers are Primbsch corrected and direct from event list***}
Count rate & $0.09 \pm 0.014$ ct s$^{-1}$ \\
Background & $\sim 0.015$ ct s$^{-1}$ \\
\tableline
\end{tabular}
\tablenotetext{a}{from the Hipparcos catalog (Perryman et al.\ 1997)}
\tablenotetext{b}{in the {\sl Einstein} energy band (0.1-4.5 keV) (Dempsey et al.\ 1993)}
\end{center}
\end{table}

%\clearpage

%\clearpage

\begin{figure}[htb!]
%\plotone{ADLeolc.eps}
\plotone{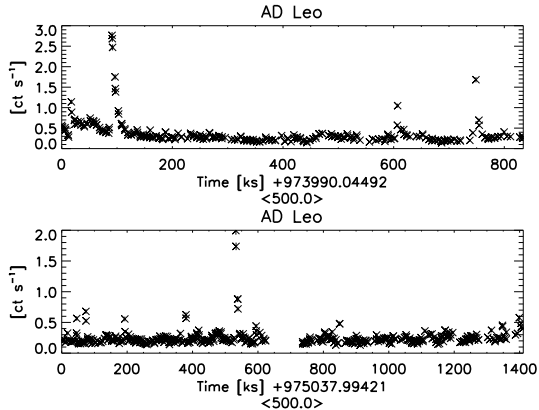}
\caption{Same as Figure~\ref{f:fkaqr_lc}, but for \adleo; upper panel
shows segments 2 \& 3 (Apr 5 - Apr 14), and the lower panel shows
segments 4 \& 5 (Apr 17 - May 4).  Many large flares are visible
(especially in segment 2, which is dominated by a large flare), as
well as significant variability in the apparently quiescent
emission.}
\label{f:adleo_lc}
\end{figure}

\paragraph{\adleo} is a well studied low-mass single flare star
(Table~\ref{t:adleo}) with a high flare rate.  A long duration exposure
was obtained by G\"{u}del et al. (2001,2002) that shows
many large flares (Figure~\ref{f:adleo_lc}).  The data were obtained
in 6 segments, and since the first segment could not be optimally reduced,
and the last segment suffered from a high background, we have ignored
them in this analysis and have concentrated on the $2^{nd} - 5^{th}$
segments.  In particular, we have carried out the analysis with the data
grouped into two sets, segments 2 and 3 forming one set and segments 4 and
5 forming the other.  The light curves show slightly different characters
in the two parts, with the former part dominated by large flares while
the latter part shows smaller identifiable flares (Figure~\ref{f:adleo_lc}).
%{***This still leaves us with 600 ks of high quality data.***}

%\clearpage
\begin{table}[htb!]
\begin{center}
\caption{{\bf \adleo} % Low-mass flare star
\label{t:adleo}}
\begin{tabular}{l l}
\tableline
\tableline
Other Names & GJ 388 / SAO 81292 \\
(RA, Dec)$_{2000}$ & (10:19:38.04, +19:52:14.2) \\
Spectral Type~$^a$ & M3V \\
Distance~$^a$ & 4.9 pc \\
$m_V$~$^a$ & $9^m.43$ \\
$B-V$~$^a$ & 1.54 \\
$L_X$~$^b$ & $8.91 \times 10^{28}$ ergs s$^{-1}$ \\
{\sl EUVE}/DS & 1999-apr-05 to 1999-apr-14 \\
\hfil & (258.6 ks) \\
\hfil & 1999-apr-17 to 1999-may-04 \\
\hfil & (347.7 ks) \\
Count rate & $0.47 \pm 0.03$ ct s$^{-1}$ [apr05-apr14] \\
\hfil & $0.31 \pm 0.03$ ct s$^{-1}$ [apr17-may04] \\
Background & $\sim 0.028-0.031$ ct s$^{-1}$ \\
\tableline
\end{tabular}
\tablenotetext{a}{from Audard et al.\ (2000)}
\tablenotetext{b}{in the {\sl Einstein} energy band (0.1-4.5 keV) (Dempsey et al.\ 1993)}
\end{center}
\end{table}

%\clearpage

%\clearpage

In our analyses (see \S\ref{s:analyz} below), we use the photon arrival
times directly, and apply the dead-time and Primbsching corrections\footnote{
Primbsching refers to the photons lost due to telemetry bandwidth, and
is measured by the ratio of the total counts incident in a quadrant of
the detector (as determined on the spacecraft; the summed counts from
all the quadrants is used to determine the instrument deadtime) to the
number of events telemetered to the ground (see the {\sl EUVE} Data
Products Guide for a detailed description of its origin and correction).}
for the particular observation to the models (see \S\ref{s:model})
over good time intervals (GTIs) defined to exclude SAA passages and
Earth blockages.  That is, we process model light curves to produce
a simulated set of photons and apply time windowing and statistical
censoring (i.e., the instrument response in the time domain) to generate
an event list that may be directly compared to the observed events.
For the sake of brevity, these corrections will henceforth
be referred to as ``instrumental'' corrections.  The source photons
are collected over a circle of radius $4''$ surrounding the source.
In all cases, $\lesssim 10\%$ of the events are estimated to be due
to background photons, except in the case of \wolf\ where it is
estimated to be 17\%.

\section{Analysis}
\label{s:analyz}

Previous attempts to determine the values of the parameters in
Equation~\ref{e:powerlaw} have concentrated on first detecting
flares in the binned light-curve and then fitting power\,law
expressions to the detected numbers.  In contrast, we {\sl assume}
the reality of a power\,law distribution and set up a {\sl model} to
compare with the data.  This model is described in \S\ref{s:model},
and the manner in which the model parameters are derived is described
in \S\ref{s:algo}.  The applicability of the method, including
verification, assumptions, advantages and disadvantages, are discussed
in \S\ref{s:apply}.

\subsection{The Model \label{s:model}}

Because flares generally
occur randomly (see \S\ref{s:assume}), we cannot directly model
the light curves, as it is not possible to ``deconvolve'' a
complex light curve by specifying the location of each flare
in a model (see, e.g., Figure~\ref{f:gj644_lc}).
Instead, rather than match the flare locations in detail, we carry out
a fitting process wherein only the number and intensity distributions
of a set of model flares are matched with the data.  This is accomplished
by comparing the distributions of photon {\sl arrival-time differences}.
The assumptions made in defining the model described below are discussed
in detail in \S\ref{s:assume}.

We first generate a set of photon arrival times by simulation from
a 3-parameter model
\begin{mathletters}
\label{e:pars}
\begin{equation}
\label{e:par1}
{\bf M}=\{\alpha,r_F,r_C\} \,,
\end{equation}
where $\alpha$
is the index of the power\,law as in Equation~\ref{e:powerlaw}, $r_F$
is the average count rate due to flares, and $r_C$ is a
constant component which is also expected to fully account for
the background (see \S\ref{s:assume}).
It is also possible to use the average {\sl total} flux
$r_T = r_F + r_C$ ($r_F < r_T$) as the defining parameter instead of $r_C$,
\begin{equation}
\label{e:par2}
{\bf M'}=\{\alpha,r_F,r_T\} \,,
\end{equation}
\end{mathletters}
with equivalent results.
The counts at time $t$ in an interval $dt$, $C(t) \sim Poisson[ r(t)\,dt ]$
are Poisson-distributed according to the instantaneous rate $r(t)$.
The rate $r(t)$ may be described as due to the sum of model counts due
to a flare component $f(t)$ and a non-flare component $r_C(t)$,
and a correction factor $\phi(t)$ that takes into account
Primbsch, dead-time, and GTIs,\footnote{
\label{f:phi}
Note that $0 \leq \phi(t) \leq 1$ determines the probability that any
photons are collected at time $t$, and in particular is identically 0
outside the GTIs.
}, i.e.,
\begin{mathletters}
\begin{equation}
\label{e:model}
C(t) \sim {\phi}(t)~Poisson[\,r_C(t)\,dt+f(t)\,dt\,] \,,
\end{equation}
where $r_C(t)$ is taken to be constant unless stated otherwise.  The
flare component $f(t)$ may in turn be written as a superposition of
numerous individual flares, i.e.,
\begin{equation}
\label{e:fcomp}
f(t) = \sum_{j=1}^{N_f} \Theta(t-t_j) F_j e^{-(t-t_j)/\tau} \,,
\end{equation}
where $\tau$ is the flare decay timescale,\footnote{
We assume $\tau$ to be fixed for purposes of simplicity.  See
\S\ref{s:assume} for a discussion of cases when $\tau$ may vary.
}
$N_f$ are the total number of flares,
$F_j$ are the peak energy intensities of individual flares
(the counts due to each flare, $c=\tau F_j$ are sampled from the
distribution represented by Equation~\ref{e:powerlaw}),
and
\begin{displaymath}
\Theta(x) = \left\{ \begin{array}{r@{\quad:\quad}l}
0 & x < 0 \\
1 & x \ge 0
\end{array} \right. %} (just to close that \left)
\end{displaymath}
is a step function to represent flare onset.

Note that not only will the placement $t_j$ and peak intensity $F_j$ of the
flares vary for each simulation, but so will the total number of
flares $N_f$.  Within the bounds of Poisson statistics, we expect
that for any given simulation, the total energy due to all the simulated
flares is determined by the average expected count rate due to flares
and the total duration of the observation, $\Delta\,T$, i.e.,
\begin{equation}
\label{e:flrct}
\int\limits_0^{\infty} dt\,f(t) =
\sum_{j=1}^{N_f} F_j\,\tau =
r_F\,\Delta\,T \,.
\end{equation}
\end{mathletters}
Note that Equation~\ref{e:powerlaw} is written as a function of
the energy deposited by a flare $E$, but assuming that the observed
counts due to this flare $c$ is proportional to $E$ (see discussion
in \S\ref{s:assume}), i.e.,
\begin{equation}
\label{e:cnorm}
\frac{dN}{dE} \propto \frac{dN}{dc} = \kappa c^{-\alpha} dc \,,
\end{equation}
we can use the model parameter $r_F$ to fix the normalization $\kappa$ of
the power\,law.  By equating the total counts due to the flare component
with the counts expected from the power\,law distribution, we get
\begin{equation}
\label{e:dndc}
r_F = \frac{\int\limits_{c_{min}}^{c_{max}} dc~c~\frac{dN}{dc}}{\Delta\,T} \,.
\end{equation}
The upper limit in the integration is defined by requiring that all the
observed counts be due to a single model flare
($c_{max}={\max\limits_j}\{F_j\,\tau\}=r_F\,\Delta\,T$), that is,
no flare model may produce a flare with more counts than are observed.
The lower limit in the integration is defined by
requiring that each flare be assigned at least 2 counts (i.e., $c_{min}=2$;
this is so that an arrival time {\sl difference} may be determined even
in the extreme case where the model may have just one weak flare -- see
\S\ref{s:algo} below).  Thus, carrying out the integral in
Equation~\ref{e:dndc} and rearranging the terms,
\begin{mathletters}
\begin{eqnarray}
\label{e:norm}
\kappa|_{(\alpha \ne 2)} &=&
\frac{ (2-\alpha) r_F \Delta\,T }
{(r_F\,\Delta\,T)^{2-\alpha} - 2^{2-\alpha}} \\
\kappa|_{(\alpha = 2)} &=&
\frac{ r_F \Delta\,T }
{{\rm ln}(r_F\,\Delta\,T/2)}
\,.
\end{eqnarray}
\end{mathletters}
For $\alpha > 2$, this implies that if $r_F \approx r_T$, then the
adopted lower limit is very close to the theoretical lower limit
to the extent of the power\,law distribution in order for it to
account for all the observed counts (see Table~\ref{t:result}).

In order to obtain the best values of the parameters that describe
the data, and a confidence range on these parameters, we carry out
a forward-fitting procedure based on a Bayesian formalism (see e.g.,
Loredo 1990 for a tutorial on the foundations of Bayesian probability
theory): we compute the probability distribution of the model
parameters given the data, which is a composite of whatever prior
information we may have on the model parameter values and the
likelihood of realizing the observed data for specified parameter
values.
That is, we derive the joint posterior probability $p({\bf M}|D,I)$
of the model parameters conditional on the data,\footnote{
The expression $p(x)$ represents the probability that the logical
statement ``$x$'' is true.  In particular, conditional statements
are written as ``$A|B$'', i.e., $p(A|B)$ represents the probability
that statement ``$A$'' is true given that statement ``$B$'' is true.
These statements may be generalized to include models and parameter
values; as an illustrative example,
$p(\{\alpha=2.1,r_F=0.1,r_T=0.3\}|D,I)=0.1$
reads as ``the probability is 0.1 that
$\alpha=2.1,~r_F=0.1,~{\rm and}~r_T=0.3$
given the data $D$ and supporting information $I$.''
}
\begin{mathletters}
\begin{eqnarray}
\label{e:bayes}
p({\bf M}|D,I) & \propto & p(\alpha|I) p(r_F|I) p(r_C|I)
\nonumber \\
 &~& \times ~ p(D|{\bf M},I) \,,
\end{eqnarray}
where $D$ represents the data, and $I$ represents assumptions necessary
to solve the problem, including the effects of instrument characteristics.
The first 3 factors on the right hand side of the equation are the {\it a
priori} probability distribution functions of the model parameters, and
the last factor, $p(D|{\bf M},I)$ is the likelihood of obtaining the
observed data given the model parameters.  Computing the posterior
probability distribution constitutes a complete solution to the
inference problem for the specified model {\bf M}.
Note that the above is a simplified form of Bayes' Theorem
wherein the
normalization factor $p(D|I)$ usually present on the right hand
side of the equation is ignored (e.g., Kashyap \& Drake 1998,
van Dyk et al.\ 2001).
If $r_T$ is used instead of $r_C$, the joint probability distribution
takes the form
\begin{eqnarray}
p({\bf M'}|D,I) & \propto & p(\alpha|I) p(r_F|r_T,I) p(r_T|I)
\nonumber \\
 &~& \times ~ p(D|{\bf M'},I) \,,
\end{eqnarray}
\end{mathletters}
with $p(r_F|r_T,I) = 0$ for $r_F > r_T$.  In the following, we make no
distinction between ${\bf M}$ and ${\bf M'}$.
The Bayesian formalism allows us to determine the probability
distributions of each of the parameters by marginalizing, i.e.,
integrating over the other parameters.  Thus we can write for
the probability distribution of $\alpha$ alone,
\begin{equation}
p(\alpha|D,I) = \int_{r_F} d\,r_F \int_{r_C} d\,r_C~~ p({\bf M}|D,I)
\end{equation}
and similarly for the other parameters (cf.\ Equation~\ref{e:palpha}).

\subsection{The Algorithm \label{s:algo}}

\begin{figure}[ht!]
%\epsscale{0.7}
%\plotone{eyefit.eps}
\plotone{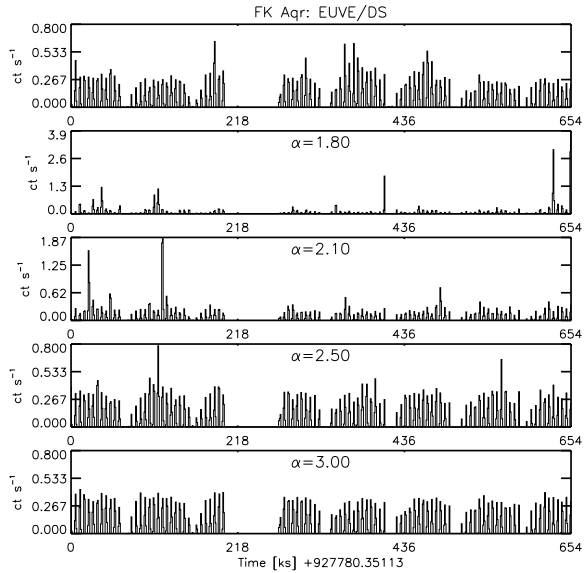}
\caption{
Illustrative comparison of data and model light curves.
The light-curve from an observation of \fkaqr\ is plotted in the
topmost panel at a bin size of 500 s.  No instrumental corrections
have been applied to the data.  In the lower panels, light-curves
derived from simulated events for various values of the index of
the power\,law distribution, $\alpha = 1.8, 2.1, 2.5, 3.0$ are
shown at the same binning, and after applying the appropriate
Primbsch and dead-time corrections and GTI filtering.  The model
curves are computed assuming $r_C=0$ in order to simplify the
comparisons.  Note the larger dynamic range in the light curves
for lower $\alpha$.  It is apparent that the light curve for
$\alpha=2.5$ is most ``similar'' to the observed light curve.
}
\label{f:eyefit}
\end{figure}

The problem facing the modeling process is illustrated in
Figure~\ref{f:eyefit}, where the light curve from the observation
of \fkaqr\ is compared with selected simulated model light
curves.\footnote{
We emphasize that these light curves are shown only for
purposes of illustration, and that the analysis does not
require binning the observed events.
}
It is easy to recognize that the model
light-curve for $\alpha=2.5$ is the most similar in character
(i.e., in the number and strength of discernible flares) to the data
light-curve, but clearly the {\sl locations} of the flares do
not match.  Normal fitting methods that rely on matching
the expected model counts in a bin with the observed counts
would fail on a problem such as this.  We thus seek to employ
a method which compares the interesting information between
the data and the model without being misled by the obvious,
though uninteresting, differences.

%\clearpage
%\clearpage

\begin{figure}[htb!]
%\epsscale{0.7}
%\plotone{dtdist.eps}
\plotone{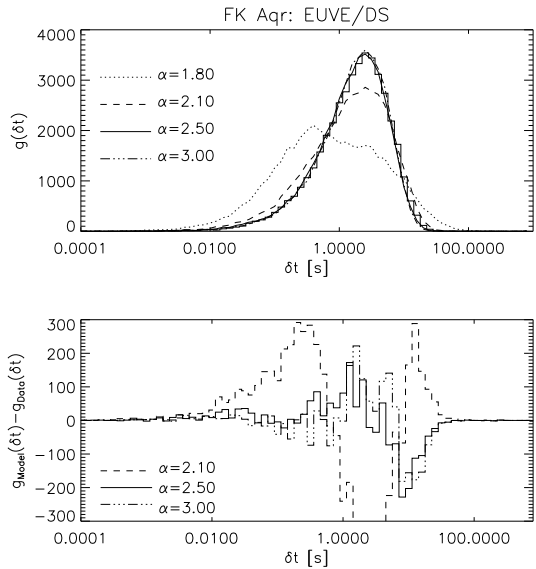}
\caption{
Comparison of distributions of arrival time differences
between the data and model.  Note that the x-axis is in log-scale.
The upper plot shows the full distributions $g(\tdel)$ for the
data (stepped curve), $\alpha=1.8$ (dotted curve), $\alpha=2.1$
(dashed curve), $\alpha=2.5$ (solid curve), and $\alpha=3.0$
(dash-dotted curve).  The models used are the same as in
Figure~\ref{f:eyefit}.  It is apparent that the first two models
are bad fits to the data.  A more detailed analysis is required to
select between the last two models.  In order to highlight the
differences in the distributions near their peaks, these differences
are shown in the lower plot, where the differences between the model
and the data distributions are plotted for $\alpha=2.1$ (dashed),
$\alpha=2.5$ (solid), and $\alpha=3.0$ (dash-dotted).  A comparison
of the $\chi^2$ values indicates that $\alpha=2.5$ is a significantly
better fit to the data than $\alpha=3$ ($\chi^2 \approx 465$ and $490$
respectively).
}
\label{f:dtdist}
\end{figure}
%\clearpage

One method that would satisfy our requirements of simultaneously
ignoring flare locations and yet be sensitive to flare numbers and
intensities is to compare the distributions of photon {\sl arrival-time
differences} $g(\tdel)$ (where $\tdel$ is the interval between consecutive
counts) between the data and the model.\footnote{
Other methods such as computing the fractal length of the events,
multi-scale analyses of light curves, etc., may also be applied
(see e.g., Vlahos et al.\ 1995).
A detailed comparison of the benefits of one method versus another
is beyond the scope of this article, but note that the results we
derive here are robust within the regime of applicability of the
adopted method (see \S\ref{s:apply}).
}
In the absence of any variability, i.e., if the light curve is
flat with expected rate $r$, the resulting set of $\tdel$ are
distributed as an exponential,
\begin{mathletters}
\begin{equation}
\label{e:dtdist}
g(\tdel) \propto r~e^{-r~\tdel} \,.
\end{equation}
In the presence of variability in the expected rate, the observed
distribution would be a superposition of many such distributions:
if the fraction of time that a source spends at
rate $r_i$ is given as $\rho_i$, then
\begin{equation}
\label{e:dtdisum}
g(\tdel) \propto \sum_i \rho_i r_i e^{-r_i \tdel} \,.
\end{equation}
\end{mathletters}
It is thus possible to distinguish between different magnitudes
of variability.
In particular, events generated from a model with
low $\alpha$ (e.g., $\alpha=1.8$, where the light curve would be dominated
by a few very large flares) would result in a distribution $g(\tdel)$
that is skewed to smaller values of $\tdel$, while those from a model
with large $\alpha$ (e.g., $\alpha=3.0$, where the light curve would
be composed of a large number of very small flares that overlap each
other so finely that it would not be possible to distinguish it from
a source with constant intensity) would approach the limiting case
of Equation~\ref{e:dtdist} above.

In Figure~\ref{f:dtdist} we show the comparison between the
distributions of arrival-time differences derived from the same
datasets in Figure~\ref{f:eyefit}.  The different curves may be
compared using any of a number of statistical methods such as
computing the $\chi^2$, applying the Kolmogorov-Smirnoff test, etc.
This approach succeeds in providing us
with an objective measure of the ``similarity'' between the
datasets; indeed of the 4 models considered in
Figures~\ref{f:eyefit} and \ref{f:dtdist}, the one with $\alpha=2.5$
has the smallest $\chi^2$ when compared with the data.  Note that
Figure~\ref{f:dtdist} also illustrates a fundamental limitation of this
method, viz., the method loses sensitivity for larger values of
$\alpha$, which may be indistinguishable from sources with a constant
intensity (see \S\ref{s:verify}).

Because the model (Equation~\ref{e:model}) is stochastic, we use
Monte-Carlo simulations to generate many realizations for each
set of model parameters.  The simulated distributions of arrival-time
differences $g_{MODEL}(\tdel)$ are compared with the corresponding
distribution derived from the data $g_{DATA}(\tdel)$ over a parameter
grid.  The likelihood is computed as
the probability density of obtaining the observed $\chi^2$ value for
N degrees of freedom (see Eadie et al.\ 1971, their Equation~4.22):
\begin{equation}
\label{e:likelihood}
p(D|{\bf M}) = \frac{ \frac{1}{2}
	\left( \frac{\chi}{2} \right)^{\frac{N}{2}-1}
	e^{-\frac{\chi}{2}} } {\Gamma(\frac{N}{2})} \,.
\end{equation}
The {\sl a priori} probability distributions for the parameters
(Equation~\ref{e:bayes}) are taken to be non-informative, and
thus flat, over the range of parameter values defining the grid.
The basic steps in the algorithm we follow are outlined below:
\begin{enumerate}
\item From the data, derive the distribution of photon arrival-time
differences, excluding the gaps in the data due to breaks in the GTIs.
\item For the specified values of model parameters ${\bf M}$
(Equation~\ref{e:pars}), obtain a realization of the photon
event-list over the duration of the observation.  This is done
by first simulating a light curve incorporating all the flare
events (Equation~\ref{e:fcomp}), added to the base emission,
and then deriving photon arrival times based on the instantaneous
count rates.
\item Because we are to compare {\sl EUVE}/DS event list data with
simulated data, instrument effects must be taken into account.  This
is encoded in the factor $\phi(t)$ (see Footnote~\ref{f:phi}).
We apply Primbsching and dead-time corrections by discarding photons
with probability $1-\phi(t)$ (i.e., sample a random number $z$ from a
uniform distribution over the interval $[0,1]$, and discard the photon
at time $t$ if $z > \phi(t)$).  The retained set of events is
identical in its instrumental characteristics to the observed data.
\item From this set of simulated photon arrival times, derive the
model distribution of arrival-time differences over a binning that
{\sl maximizes} the reduced $\chi^2$.\footnote{
We do not know the optimal binning for $g(\tdel)$ {\it a priori}.
The binning must be chosen such that differences between the two
distributions being compared are highlighted to best advantage,
at a scale that is determined by the datasets themselves.  Because
the total number of photons in the datasets being compared are
approximately the same, very coarse binnings and very fine binnings
both produce low values of the reduced $\chi^2$, and we adopt as
the optimal binning that which maximizes this value.
}
\item Compute the likelihood as in Equation~\ref{e:likelihood} and
the {\sl a posteriori} probability at the specified grid point as
in Equation~\ref{e:bayes}.  Note that the number thus obtained is
not normalized, and so must not be used to compare, for example,
the relative probabilities of different types of models.
\end{enumerate}

In practice, the above algorithm must be enhanced by 
some additional steps.  For instance, the likelihood may be
artificially reduced if a large model flare is fortuitously located
coincident with large dead-time.  We therefore shift the Primbsching
and dead-time corrections by a random interval and recompute the
arrival-times.  This is mathematically equivalent to shifting the
simulated events, but is done in this fashion because of the lower
computational cost.  Typically 3 such shifts are carried out for each
simulated light curve, and the best comparison is chosen.  In addition,
in order to derive a robust estimator, we perform $\approx 15-20$
simulations for each ${\bf M}$ (or ${\bf M'}$) and adopt the median
value of the set as the final value of $p({\bf M}|D,I)$.
The grid of model parameters are chosen such that $\alpha$ is explored
over the useful range of the algorithm ($1.5{\ldots}3.0$; see
\S\ref{s:verify}); $r_T$ in a range within 3$\sigma$ of the average
count rate; and $r_C$ and $r_F$ ranges from $\approx 0$ to the average
count rate.
There are typically $\sim 15$ grid points for each parameter.

The probability distribution along any of the axes is then obtained
by summing over the other axes and normalizing, e.g.,
\begin{equation}
\label{e:palpha}
p(\alpha|D,I) =
\frac{ \sum\limits_{r_F} \sum\limits_{r_C} p({\bf M}|D,I) }
{\sum\limits_{\alpha} \sum\limits_{r_F} \sum\limits_{r_C}
p({\bf M}|D,I) } \,.
\end{equation}
The derived probability distributions may be summarized by their
means and variances,\footnote{\label{f:map}
Other well-known methods of summarizing probability distributions
include noting the MAP (maximum {\it a posteriori} value; the mode of
the distribution), as e.g.,
$p(\alpha_{_{MAP}}|D,I) = \max\limits_\alpha \{ p(\alpha|D,I) \}$,
or a range corresponding to an integrated area under the curve,
equivalent to some defined probability
$\pi = \int\limits_{\alpha_{min}}^{\alpha_{max}} p(\alpha|D,I)$,
such that $\alpha_{_{MAP}} \in [\alpha_{min},\alpha_{max}]$,
etc.  Unless otherwise specified, we always report the mean values
and $1\sigma$ errors.
} e.g.,
\begin{mathletters}
\begin{eqnarray}
\label{e:varmean}
\overline{\alpha} = \frac{ \sum\limits_\alpha ~ \alpha ~ p(\alpha|D,I) }
{\sum\limits_\alpha ~ p(\alpha|D,I)} \,, \\
var(\alpha) = \overline{\alpha^2} - (\overline{\alpha})^2 \,.
\end{eqnarray}
\end{mathletters}

\subsection{Applicability}
\label{s:apply}

As was demonstrated above
(\S\ref{s:algo}, Figures~\ref{f:eyefit}\,\&\,\ref{f:dtdist}),
the distributions of arrival-time differences $g(\tdel)$
provide an objective means to compare event lists that
are dominated by stochastic events.
Here, we verify that the algorithm gives reasonable
results by creating simulated datasets and obtaining
best-fit values for them (\S\ref{s:verify}), then
discuss the effects of some of the assumptions we have
made in formulating the problem (\S\ref{s:assume}), and
detail the advantages and disadvantages of the adopted
method (\S\ref{s:advantage}).

\subsubsection{Verification}
\label{s:verify}

\begin{figure}[htb!]
%\plotone{verify_alpha.eps}
\plotone{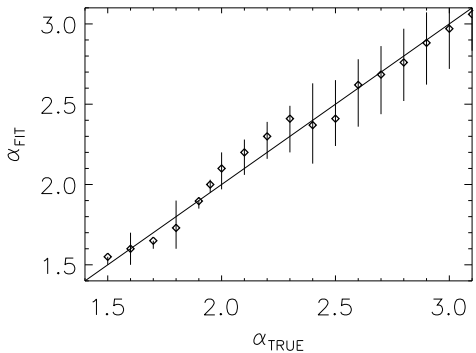}
\caption{Verifying the method.  The best-fit values $\alpha_{_{FIT}}$
obtained for simulated datasets generated using a specific values
$\alpha_{_{TRUE}}$ are shown as diamonds, and the 90\% credible regions
are shown as vertical bars.  The line of equality (solid line) is
also shown.}
\label{f:verify}
\end{figure}

We show here that the algorithm works as expected by simulating
well defined datasets and then ``fitting'' to them.
We generated flare models with specific values of $\alpha$
(labeled $\alpha_{_{TRUE}}$)
ranging from 1.5 to 3, and for the sake of simplicity, held
$r_F=0.5$ ct s$^{-1}$ and $r_C=0.1$ ct s$^{-1}$.  Simulations
were carried out over a time interval $\Delta\,T=100$ ks and assuming
a fixed flare decay timescale of $\tau=3$ ks.  These values were
chosen as being typical of {\sl EUVE}/DS observations.  The datasets to
be fit to were chosen by first simulating 9 separate event lists
and then choosing the one with the median number of counts, in order
to avoid contaminating the verification process with extremal datasets.
The fits were carried out as described in \S\ref{s:algo}, and the
resulting best-fit values $\alpha_{_{FIT}}$ are shown in
Figure~\ref{f:verify}.  The fitted values follow the true values
closely, for all 3 parameters.

%\clearpage
%\clearpage

The method works well over the range of interesting $\alpha$,
that is, spanning the value $\alpha=2$, and is capable of
distinguishing between sources with $\alpha$ above or below
this critical value (see \S\ref{s:intro}).  The sensitivity
of the method naturally decreases as $\alpha$ increases to
$\gtrsim 3$ when the datasets become indistinguishable from
that of a steady source.  However, this is not a hard limit
and can be extended for datasets with larger average flare
component intensity $r_F$ and longer observations $\Delta\,T$.
The reliability also decreases as $\alpha$ decreases to
$\lesssim 1.5$ since at these values of $\alpha$ the simulations
are dominated by a very few but very large flares and are
therefore subject to large fluctuations, and hence are not robust.
A larger number of simulations at each grid point becomes
necessary when a dataset exhibits smaller values of $\alpha$.

\subsubsection{Assumptions}
\label{s:assume}

We have made a number of simplifying assumptions in our analysis,
and these are discussed below, with particular attention to the
effect they have towards the robustness of the results.  In general,
all our assumptions are conservative, in the sense that they all
act to generate a best-fit $\alpha$ and $r_F$ that is smaller than
the true $\alpha$ and $r_F$.  That is, the main results expounded
here, that the flare distributions on active stars have $\alpha>2$,
and that the apparently quiescent emission is dominated by flares,
are not affected.

\paragraph{Power\,laws:}
Energy release due to flare events in both solar and stellar
environments has been well established to be highly intermittent
and that the events are distributed as
power\,laws spanning many decades in energy (e.g., Crosby et al.\
1993, G\"{u}del et al.\ 1997, G\"{u}del 1997, Krucker \& Benz 1998,
Osten \& Brown 1999, Audard et al.\ 2000, Aschwanden et al.\ 2000,
Veronig et al.\ 2002).
This suggests the absence of a characteristic scale for the intensity
of a flare event, and is understood to arise from
avalanche or SOC (self-organized critical) models (Lu \& Hamilton 1991,
Vlahos et al.\ 1995, Georgoulis \& Vlahos 1998, Krasnoselskikh et
al.\ 2001).  Nevertheless, there is evidence from studies of solar
flares that the index of the power\,law distribution does not remain
the same in the microflare and nanoflare range (cf.\ Aschwanden et
al.\ 2000 [their Figure~10], Winebarger et al.\ 2001).  Studies
of stellar flares (e.g., Ambruster et al.\ 1987) also suggest that
more than one type of plasma instability may be present, and that
the distribution may steepen or change at lower flare energies;
for instance,
compare $\alpha \approx 1.8$ found by Osten \& Brown (1999) with
the generally larger values found by Audard et al.\ (2000) who
include much weaker flares in their analysis (also see the
results from \adleo\ presented here in \S\ref{s:result}).
Studies of flare distributions arising from the transition region
also show a similar dichotomy (Robinson et al.\ 2001).
Such changes in power\,law indices could arise from
a variable driving mechanism (Georgoulis \& Vlahos 1998).
It is therefore possible that the true distribution departs from exact
self-similarity
in some complicated manner.  However, present data, especially in the
case of stellar flares, are insufficient to detect these departures
(Audard et al.\ 1999,2000; also G\"{u}del et al.\ 2001,2002).
Here we assume that a
single power\,law index is valid over at least 4 orders of magnitude
(see Table~\ref{t:result}) in flare energy.  If the distribution
does steepen for lower flare intensities, then note that first,
the steepest parts, which approach the limiting case of constant
emission (see Equation~\ref{e:dtdist}), will contribute to enhancing
the constant intensity $r_C$ (as described in \S\ref{s:algo}).
Hence the fitted
$r_F$ will be a lower limit to the true value.  Second, the fitted
$\alpha$ will be a weighted average biased towards the high
count rate, shallower distribution, and inasmuch as a ``true'' value
of $\alpha$ may be said to exist, it would be greater than the fitted
value.

\paragraph{Decay Timescales:}
We analyze the data by fitting the 3 parameters $\alpha, r_F,$ and $r_C$
of the model (Equation~\ref{e:pars}).  We assume that the flare decay
timescale $\tau$ (Equation~\ref{e:fcomp}) is fixed (usually at 3 ks, as
suggested by the detectable flares, and the radiative cooling timescales
suggested by {\sl ROSAT} observations; see e.g., Giampapa et al.\ 1996)
and is the same for all flares.
However, it is well known that
$\tau$ varies for individual flares on active stars (e.g., Pallavicini
et al.\ 1990) and especially so for RS\,CVn stars (Osten \& Brown 1999).
Note though that we confine ourselves to a specific class of active stars
-- low-mass main-sequence stars that flare frequently -- and exclude
RS\,CVn stars from our sample.
There is evidence that flare duration scales with flare energy
in various passbands
(e.g., Crosby, Aschwanden, \& Dennis 1993; also,
Vlahos et al.\ 1995 for avalanche models,
Temmer et al.\ 2001 for H$_\alpha$ flares,
Georgoulis, Vilmer, \& Crosby 2001 for deca-keV flares,
Robinson et al.\ 2001 for chromospheric and transition region events,
Veronig et al.\ for GOES SXR flares)
such that more intense flares appear to last longer.
Based on avalanche
model simulations, Lu et al.\ (1993) find that on average,
an event with decay timescale $\tau$ corresponds to an
energy release $E \propto \tau^{1.77}$.
(See also the discussion below about sympathetic flaring,
which could affect measurements of flare durations.)
But note that the decay timescales for soft X-ray flares (such as
the ones we are concerned with) are primarily dependent on flare
temperature and plasma density, and secondary effects such as changes
in the heating rate come into play only for very large flares where
the heating timescales approach or surpass the radiative cooling
timescales.
In a model that incorporates such variations of $\tau$, the fainter
flares will have larger peak rates than in the regular model (in
order to have the same total energy output), and the distribution
of arrival-times $g(\tdel)$ will be skewed towards smaller $\tdel$.
Thus, fitting to data that may be generated in this manner using a
$\tau$ fixed by the higher intensity flares would result in
$\alpha_{_{FIT}} < \alpha_{_{TRUE}}$ (see Figure~\ref{f:dtdist}) and
therefore the fitted values would be lower limits to the true
indices.  Conversely, fitting a model where $\tau$ decreases with flare
energy to a dataset that is not generated in this manner will result
in $\alpha_{_{FIT}} > \alpha_{_{TRUE}}$.
(See also discussion by G\"{u}del et al.\ 2001,2002.)
In order to test the sensitivity of our datasets to these variations in
modeling, we
have carried out fits to the data using models with an energy-dependent
timescale, $\tau \propto E^{\beta}$, in particular, $\beta=\frac{1}{4}$
(G\"{u}del et al.\ 2002, based on fits to {\sl EXOSAT}
flare decay timescales of Pallavicini et al.\ 1990).
We find that the
best-fit value of $\alpha$ {\sl increases} by $\sim 0.3-0.4$ for the
complex model, e.g., for segment 3 of \adleo\ $\alpha$ changes from
2.2 to 2.7 .
We thus conclude that using the simpler model ($\tau=constant$) is
preferable in that we do not overestimate $\alpha$, and further note
that stronger dependences, e.g., that suggested by the theoretical
study of Lu et al.\ (1993; $\tau \propto E^{0.56}$) would result in
even larger fitted values of $\alpha$.
In addition, we have also explored the sensitivity of fits to the
adopted value of $\tau$, by generating simulated data with small
decay timescales $\tau_{_{SIM}}$ and then fitting to it a model
with a larger decay timescale $\tau_{_{FIT}}$.  We find that
as expected the best-fit $\alpha$ is {\sl smaller} than the true
value, i.e.,
$\alpha_{_{FIT}}|_{(\tau_{_{FIT}}>\tau_{_{SIM}})}~<~\alpha_{_{SIM}}|_{\tau_{_{SIM}}}$.\footnote{
This effect is large only for small values of $\alpha$.  For example,
$\{\alpha_{_{SIM}}=1.7; \tau_{_{SIM}}=1000 \}$ is fit by
$\{\alpha_{_{FIT}}=1.5 \pm 0.02 ; \tau_{_{SIM}}=3000\}$,
whereas $\{2.1 ; 500\}_{_{SIM}}$ is fit by
$\{2.16 \pm 0.17 ; 3000\}_{_{FIT}}$ (i.e., is indistinguishable from
the true value), etc.
Note that, as expected, this exercise also shows that when the data
are dominated by multiple overlapping flare events, the results are
insensitive to the precise value of the decay timescale, and effects
such as stellar rotation may be ignored.
}
The reverse holds true for the opposite case, i.e.,
$\alpha_{_{FIT}}|_{(\tau_{_{FIT}}<\tau_{_{SIM}})}~>~\alpha_{_{SIM}}|_{\tau_{_{SIM}}}$,
in which case the best-fit values imply flare distributions that are
steeper than they should be.  Thus, the relatively large value we
adopt for $\tau$ (3 ks) provides a conservative estimate, and the
fitted values may be considered lower limits to the true values.

\paragraph{Rise Times:}
We model the flares as having an instantaneous rise and a slow
exponential decay (Equation~\ref{e:fcomp}).  In general this is
a good approximation, since the rise times are short compared to
the flare duration (e.g., Reale, Peres, \& Orlando 2001, Temmer et
al.\ 2001).  Further, our algorithm is insensitive to the exact
sequence of the emission intensities (i.e., the process of
forming $g(\tdel)$ destroys the sequential information in the
event lists), and thus any discernible flare rise times will be
indistinguishable in their effects from flares of small decay
times.  Thus, this approximation reduces to the problem
discussed above, that of a sequence of flares that are
``contaminated'' by flares with smaller decay timescales, with
similar effects on $\alpha$ and $r_F$.

\paragraph{Flare Waiting-times:}
Much work has been done to characterize the time between flares
within individual active regions.  If we consider each flare to
be an independent event then the waiting time between flares
is a Poisson process (Rosner \& Vaiana 1978).  However, actual
observations of solar Hard X-ray flares show that the waiting-time
distribution (WTD) is a power\,law in intervals, with index ranging
from $-2.16$ to $-2.4$ (Boffetta et al.\ 1999, Wheatland 2000,
Lepreti, Carbone, \& Veltri 2001), though Moon et al.\ (2001) find
that for strong solar flares (strength greater than C1) the waiting
times are indeed well characterized by a Poisson distribution.  
The power\,law distribution has been interpreted as due to
sympathetic flaring (i.e., a cascade of small flares that follow
a large flare, thus invalidating the assumption of event independence)
by Wheatland, Sturrock, \& McTiernan (1998), and adapted within SOC
models as a non-stationary random process (Norman et al.\ 2001).
However, since we can only observe disk-integrated flux in stellar
data (it is not possible to monitor individual active regions),
event independence is a better approximation.  We thus assume in
our modeling that the stellar flare WTD is Poisson.  In such a case,
the model undercounts the number of flares separated by short
intervals (Wheatland et al.\ 1998).  These flares would generally
be associated with the stronger flares that set off a cascade, thus
effectively increasing the decay timescale for large flares.  As
argued above, this causes $\alpha_{_{FIT}} < \alpha_{_{TRUE}}$.

\paragraph{Energy Deposition:}
We have implicitly assumed that the observed counts track the
energy deposited by the flares linearly (see
Equations~\ref{e:powerlaw} and \ref{e:cnorm}).
That is, we assume that the energy deposition process that
causes the flare event (generally considered to be magnetic
reconnection -- see Parker 1988) is, first, sufficiently
energetic that a large fraction of the deposited energy goes
towards thermal loading of the plasma and not into bulk motions
(see Winebarger et al.\ 2001); second, that the resulting
plasma temperatures for all flare energies lie near $10^7$ K;
and third, that the temperature evolution of the plasma after
the flare event is not drastic.
These assumptions are supported by the emission measure analysis
of several {\sl Yohkoh} flares by Reale et al.\ (2001), who
find that solar flares are dominated by emission at $\sim 10$ MK.
Hydrodynamic modeling of an ensemble of flaring loops by
G\"{u}del et al.\ (1997) and G\"{u}del (1997) also shows that
the bulk of the flare DEM lies above 6 MK; indeed G\"{u}del (1997)
finds that the DEM is bimodal around 10 MK, attributable to the
slightly smaller temperatures generated by the weaker flares
(but which are still significantly hotter than the temperatures
achieved by the quiet Sun).  Thus, while it is reasonable to
expect that flare emission would evolve from temperatures of
$\sim 20$ MK to $\sim 5$ MK, and that flare events that are
less energetic would heat the plasma to a lower temperature,
in practice the effects of such variations are very little.
Furthermore, because we model the distribution in counts space,
i.e., $\frac{dN}{dc}$ rather than $\frac{dN}{dE}$ directly,
we invariably obtain distributions that are {\sl shallower}
than the true distributions\footnote{
Suppose we write the observed counts generated in a detector $c_{obs}$
due to emitted energy $E_{true}$ as a power\,law that deviates from
linearity by a small amount, $c_{obs} \propto E_{true}^{1+\delta}$,
$\delta>0$.  That is, as $E_{true}$ decreases, the counts produced
in the detector decrease at a faster than linear rate, as would
be expected to happen if the lower-energy flares result in lower
plasma temperatures, and the detector has a smaller response to
these temperatures.  This is a reasonable
approximation for broad-band instruments such as the {\sl EUVE}/DS,
though it may not be applicable for small-passband detectors such
as {\sl TRACE} (see Aschwanden \& Charbonneau 2002).  If
$\frac{dN}{dE_{true}} \propto E_{true}^{-\alpha_{true}}$, then
$\frac{dN}{dc_{obs}} \propto c_{obs}^{-\alpha_{obs}}$, where
$\alpha_{obs}=\frac{\alpha_{true}+\delta}{1+\delta}$, and is always
smaller than $\alpha_{true}$ for $\delta>0$ and $\alpha_{true}>1$.
Conversely, if $\delta<0$, as may happen for extremely high
flare energies and temperatures that lie above the best response
of the detector, the observed distribution will be steeper than
the true distribution, $\alpha_{obs} > \alpha_{true}$; however
this case has little effect on our analysis results because of
the wide temperature response of the {\sl EUVE}/DS, and any
flares that fall outside its range would be too few in number to
affect the results.
}
(see also extensive discussion in G\"{u}del et al.\ 2002).
In the {\sl EUVE}/DS, changes in temperature of this magnitude
causes the observed counts to vary by a factor of $\approx 2$.
Furthermore, the range of flare energies we can model (see
Table~\ref{t:result}) are much larger than the relatively low-energy
``explosive events'' that are characterized by large non-thermal
velocities (Winebarger et al.\ 2001 and references therein), and
which may not contribute significantly to the plasma heating.
That is, we assume that lower energy depositions that
may heat the plasma to lower temperatures would not be detectable
by the {\sl EUVE}/DS in any case.
Note that the assumed abundances will also affect this to some
extent, but its effect is minimal for a broad-band instrument
such as is used here.  We therefore assume that an observed
{\sl EUVE}/DS count corresponds to a photon of average energy
$1.7 \times 10^{-11}$ ergs cm$^{-2}$ ct$^{-1}$ over the 59-250 \AA\
range produced by a plasma at $10^7$ K (PIMMS v2.5).

\paragraph{Cut-offs:}
We have adopted upper and lower limits to the extent of the
power\,law distribution (see Equation~\ref{e:dndc}) that are
based strictly on numerical expediency: the upper limit is
set by the requirement that the largest possible flare can
produce no more than the observed number of counts, and the
lower limit is defined by the minimum number of counts needed
to define an interval.  How do these limits compare with
theoretical expectations?  First note that the solar flare
distribution has been explored to much lower energies,
$E \sim 10^{23}$ ergs (Winebarger et al.\ 2001), than has
been achieved for stellar observations.
Recently Katsukawa \& Tsuneta (2001) have estimated that
the probable energy of Parker-type nanoflares is $<10^{22}$ ergs.
In contrast, we find that if the observed {\sl EUVE} emission
is assumed to originate entirely in flares, then it is
sufficient to extend the power\,law distributions to energies
$E \sim 10^{28-29}$ ergs (see Table~\ref{t:result}).  We
suggest that the discrepancy is not due to a higher cut-off
energy in the stellar case, but rather that our analysis is
physically limited due to a combination of the lack of
instrument sensitivity, possible power\,law changes, and
systematic bias due to temperature effects in very small flares
(see above).  By modeling flares as driven dissipative
avalanche systems, Lu et al.\ (1993) predict a high-energy rollover
in the energy distribution whose magnitude depends on the size of
the active region where the flares occur.  Kucera et al.\ (1997)
have applied this to {\sl SMM}/HXRBS data and find empirically
that the total energy of an event has a maximum, $E_{cutoff} \approx
5 \times 10^{28} A_{\rm{\mu}hs}^{5/4}$ ergs, where $A_{\rm{\mu}hs}$
is the total area of active regions in units of solar micro-hemispheres,
i.e., $1 {\rm{\mu}hs} = 3.04 \times 10^{18}$ cm$^{2}$.  Thus, for
active stars such as the ones we are considering, where
active regions covering large fractions of the stellar surface
are expected ($A_{\rm{\mu}hs} >> 10^4$), total flare energies
in excess of $10^{33}$ ergs are achievable, and thus our
analyses are valid up to these energies.

\paragraph{Background Corrections:}
All the stars in our sample are strong sources of EUV
emission, and the background is generally small compared to
the source strength.  We do not subtract the background
from the datasets, nor model it separately, but assume that
the constant component in the flare model (Equation~\ref{e:model})
includes contamination due to the background.  That is, we
assume that the background does not vary on timescales smaller
than the adopted decay timescale, and that any variations that
exist in the instrumental and astrophysical background are small
in magnitude and do not contribute to the flare component.  Any
departures from strict constancy in the background will result
in a poorer determination of $r_C$ because of the larger spread
in the distribution of arrival times, $g(\tdel)$ (see
Figure~\ref{f:dtdist}).  In cases where the above assumptions
are invalid, the background data will contaminate the signal,
but as estimated in Tables~\ref{t:fkaqr}-\ref{t:adleo}, this
contamination will be $\lesssim 10\%$.

\subsubsection{Advantages and Disadvantages \label{s:advantage}}

Unlike previous methods that determine flare number
distributions non-parametrically, that is, by directly
counting the number of detected flares, we {\sl model} the
flare distribution as a power\,law, and generate photon
arrival-times to compare with the observed event lists.
Thus the validity of our results is directly dependent on
the applicability of the model (see extensive discussion
above in \S\ref{s:assume}), i.e., the derived parameter
values are only as good as the model.  Thus, our results
are correct subject to the caveat that we require that all
flares are physically ``similar'' and that the flare
distribution follows a power\,law.  But the fact that we
model the distribution means that we can usefully extend
the analysis to the regime where the light curves are
dominated by large numbers of small, undetectable flares,
and hence this method is best suited to study continuous
micro-flaring.

Because the model is stochastic in nature, any realization of
the photon arrival times can be significantly different from
the observed event list even if the parameter set matches
exactly.  Thus a large number of simulations must be carried
out for each set of parameters, which is a time consuming
process.  On the other hand, the stochastic nature of the
model renders unnecessary an exact match between the model
realization and the observed light curve, allowing us to
explore weak flare events.

As shown in \S\ref{s:algo} (Figure~\ref{f:dtdist}) and
\S\ref{s:verify}, the method loses sensitivity for $\alpha
\gtrsim 3$ and loses stability for $\alpha \lesssim 1.5$.
However, over the range of $\alpha$ that is of scientific
interest, i.e., spanning the critical value $\alpha=2$, the
method is robust and can discriminate between the two cases
where flare emission {sl may be} a significant contributor to
the coronal emission budget ($\alpha > 2$), or where the
light curve may be dominated by large flares which do not
contribute significantly to the energy budget
($\alpha<2$).\footnote{Note
that the mere fact that $\alpha>2$ does not guarantee that the
coronal emission is dominated by flaring, but the normalization
of the power\,law distribution must also be large.  Our approach allows
us to independently determine the flare contribution $r_F$ to the
total count rate.  The results (see \S\ref{s:result}) show that the
normalization is such that the flare contribution is generally $>50\%$.}

\section{Results}
\label{s:result}

We have applied the algorithm detailed above (\S\ref{s:analyz})
to {\sl EUVE}/DS data on \fkaqr, \wolf, and \adleo\ (see \S\ref{s:data}).
The results are summarized in Table~\ref{t:result}.  Below we comment
on each dataset in detail.

%\clearpage
\begin{table*}[htb!]
\begin{center}
\caption{Summary of Results \label{t:result}}
\begin{tabular}{l c c c c c}
\tableline
\tableline
Star &
\multicolumn{1}{c}{$\overline{\alpha}$\tablenotemark{a}} &
	\multicolumn{1}{c}{Flares \%\tablenotemark{b}} & 
	\multicolumn{1}{c}{$\kappa$\tablenotemark{c}} &
	\multicolumn{1}{c}{$E_{range}$\tablenotemark{d}} & 
	\multicolumn{1}{c}{$E_{min}$\tablenotemark{e}} \\
\hfil & \hfil & \hfil & \hfil &
\multicolumn{1}{c}{log$_{10}$[ergs s$^{-1}$]} &
\multicolumn{1}{c}{[$\times 10^{28}$ erg]} \\
\tableline
\fkaqr & $2.60 \pm 0.34$ & $\sim$ 50 (65) & $2.09\times 10^4$ & 29.48 - 33.54 & 9.87 \\
\wolf & $2.74 \pm 0.35$ & 70 (85) & $8.6\times 10^3$ & 29.12 - 32.67 & 8.64 \\
\adleo & \hfil & \hfil & \hfil & \hfil & \hfil \\
\hfil $[2,3]$ & $2.17 \pm 0.03$ & 80 (80) & $2.12\times 10^4$ & 28.99 - 33.66 & 8.20 \\
\hfil $[4,5]$ & $2.32 \pm 0.11$ & 65 (75) & $2.8\times 10^4$ & 28.99 - 33.52 & 3.40 \\
\tableline
\end{tabular}
\tablenotetext{a}{Power\,law index.}
\tablenotetext{b}{For the average rate, $\overline{r_F}$ as a
percentage of the average total rate; within brackets, for the
mode $r_{F_{_{MAP}}}$.}
\tablenotetext{c}{Normalization factor for power\,law distribution, for
best-fit parameters (see Equation~\ref{e:norm}).}
\tablenotetext{d}{Range of flare energies included in calculations
(see Equation~\ref{e:dndc}).}
\tablenotetext{e}{Minimum flare energy to which power\,law should be extended
in order for the flare component to account for the entire emission.
}
\end{center}
\end{table*}
%\clearpage

\subsection{\fkaqr}

As anticipated above (see Figure~\ref{f:dtdist}), the
maximum {\it a posteriori} value (MAP; see footnote~\ref{f:map})
for \fkaqr\ are
$\alpha_{_{MAP}} \approx 2.5$,
$r_{F_{_{MAP}}} \approx 0.22$ and
$r_{T_{_{MAP}}} \approx 0.36$,
corresponding to a flare contribution of $\approx 65\%$ to the total
emission.  The joint probability distribution of $\alpha$ and $r_F$,
marginalized over $r_T$, is shown in Figure~\ref{f:fkaqr_jprob}, and
the individual probability distributions of $\alpha$ and $r_F$,
marginalized over the other parameters, are shown in
Figure~\ref{f:fkaqr_prob}.
The best-fit values are
$\alpha = 2.60 \pm 0.34$,
$r_F = 0.19 \pm 0.12$, and
$r_T = 0.37 \pm 0.01$.
The correlations between the parameters
are evident in $p(\alpha,r_F)$.
Note that there is a small probability ($\sim 0.1$) that
\{$r_F<<r_T$ \& $\alpha<2$\}.
It is worth pointing out the cause of the large values of the
probability for small $r_F$.
First note the presence of a secondary peak at
($\alpha=2.2$,~$r_F=0.05$) which contributes to a larger
spread in the uncertainty in $r_F$, and incidentally also
indicates the potential existence of multiple power-law
components.
Second, this effect is a measure of the stability of the
model best-fit parameters to the data; if there are large
parts of the parameter space which provide adequate (though
not good) fits to the data, their contributions are enhanced
due to the larger volume of the space they occupy.  This
is indeed the case here for combinations of small values
of $r_F$ and $\alpha$, where the skewness induced in
$g_{MODEL}(\tdel)$ by $\alpha < 2.5$ is minimized due to
the lower weight accorded to it because of the smaller
values of $\kappa$.
Data from {\sl Chandra} or {\sl XMM-Newton},
with their higher expected count rates, are necessary to explore
the region of smaller values of $\tdel$ and thus better constrain
the parameter ranges.

%\clearpage
\begin{mathletters}
\begin{figure}[htb!]
%\plotone{FKAqr_cont.eps}
\plotone{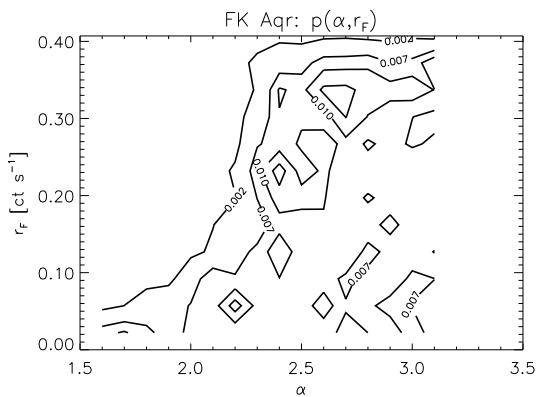}
\caption{Joint posterior probability distribution of $\alpha$ and $r_F$,
marginalized over $r_T$, for \fkaqr.  The peak of the distribution lies
at $\alpha_{_{MAP}}=2.5$ and $r_{F_{_{MAP}}}=0.22$.}
\label{f:fkaqr_jprob}
\end{figure}
\end{mathletters}

%\clearpage

\begin{mathletters}
\begin{figure}[htb!]
%\plotone{FKAqr_af.eps}
\plotone{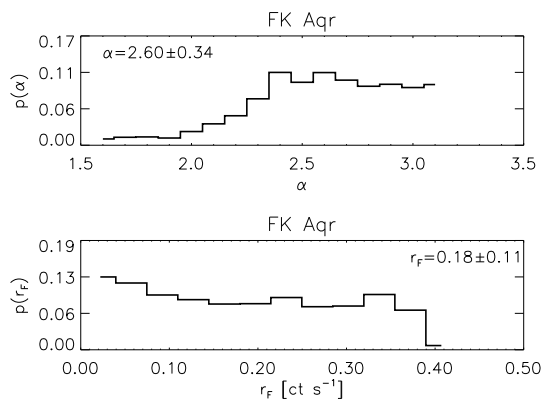}
\caption{Marginalized 1-D posterior probability distributions of
$\alpha$ and $r_F$ for \fkaqr.  Note the apparently large contribution
at small $r_F$, which is a consequence of the fact that the flare
contribution is poorly determined but cannot be ruled out for any $\alpha$
for small values of $r_F$.
}
\label{f:fkaqr_prob}
\end{figure}
\end{mathletters}
%\clearpage

\subsection{\wolf}

The light curve of \wolf\ (Figure~\ref{f:gj644_lc}) shows considerable
variability with some relatively weak flare-like events.  This is a
signature of a flare distribution with large values of $\alpha$, and
indeed detailed analysis confirms this impression; we find
$\overline{\alpha}=2.74\pm0.35$, with an upper bound that is not well
constrained
(Figure~\ref{f:gj644_prob}).  From the joint probability distribution
$p(\alpha,r_F)$, the most probable values are $\alpha_{_{MAP}} \approx 2.5$
and $r_{F_{_{MAP}}} \approx 0.065$, suggesting that it is likely that almost
all the observed emission originates in flare-like events.

%\clearpage
\begin{figure}[htb!]
%\plotone{V1054Oph_cont.eps}
\plotone{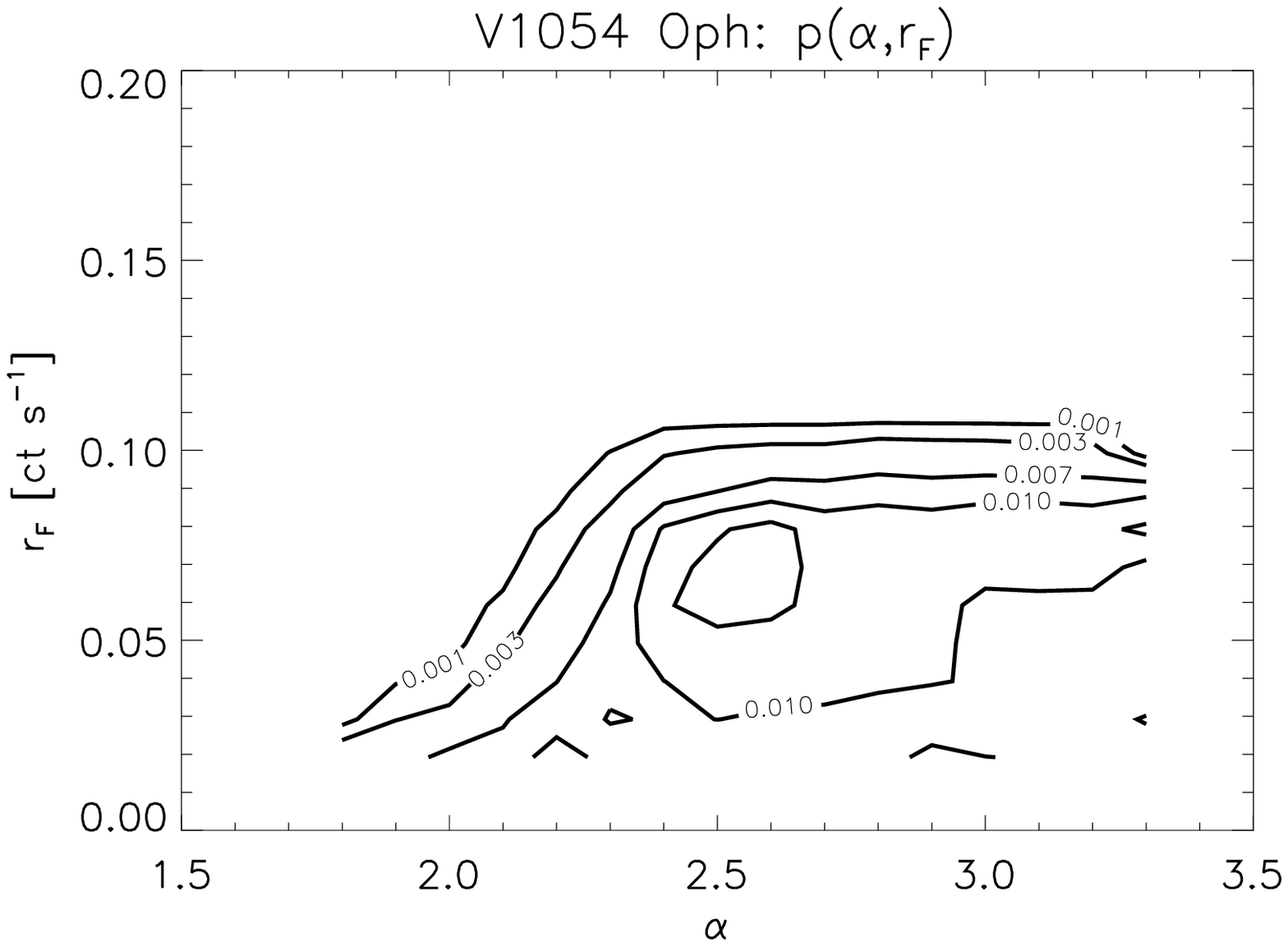}
\caption{As Figure~\ref{f:fkaqr_jprob}, for \wolf.
}
\label{f:gj644_jprob}
\end{figure}

%\clearpage

\begin{figure}[htb!]
%\plotone{V1054Oph_af.eps}
\plotone{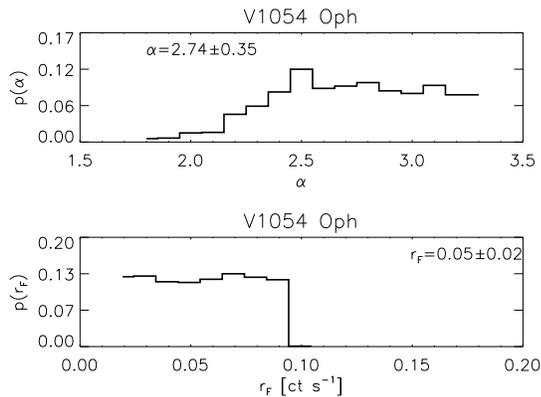}
\caption{As Figure~\ref{f:fkaqr_prob}, for \wolf.  Note that there
is significant mass in the distribution $p(\alpha)$ for large values
of $\alpha$, which implies that the upper-bound is not well constrained.
However, the lower-bound is well determined to be $\gtrsim 2.2$.
}
\label{f:gj644_prob}
\end{figure}
%\clearpage

\subsection{\adleo}

\begin{figure}[htb!]
%\plotone{ADLeo23_af.eps}
\plotone{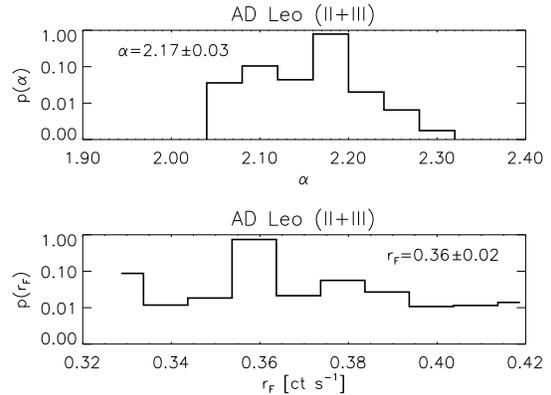}
\caption{As Figure~\ref{f:fkaqr_prob}, for segments 2 and 3 of \adleo.
The parameters are well determined due to the large size of the dataset.
Values of $\alpha<2$ can be emphatically ruled out.
}
\label{f:adleo_prob_23}
\end{figure}

\begin{figure}[htb!]
%\plotone{ADLeo45_af.eps}
\plotone{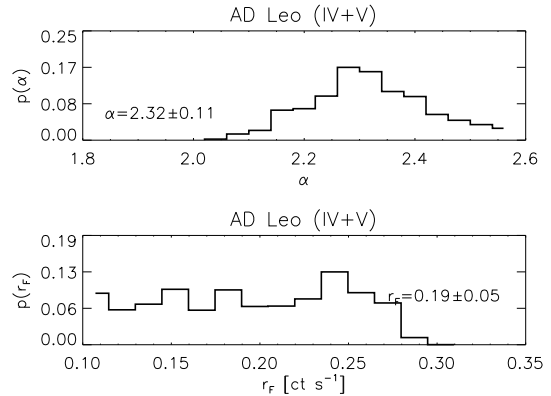}
\caption{As Figure~\ref{f:adleo_prob_23}, for segments 4 and 5 of \adleo.
The best-fit value of $\alpha$ is larger than for segments 2 and 3,
though the probability distributions do overlap.  The relatively large
spread in $r_F$ is due to the larger allowed spread in $\alpha$
compared to the earlier data.  The most probable values are
$\alpha_{_{MAP}}=2.3$ and
$r_{F_{_{MAP}}}=0.24$,
which implies that the data are almost entirely due to flaring emission.
}
\label{f:adleo_prob_45}
\end{figure}

In general, the \adleo\ data have been analyzed in two separate batches
because of the large time intervals and large number of counts involved
(which leads to very long computation times) and also because the
character of the light curve
(see Figure~\ref{f:adleo_lc}) appears to change from being dominated by
an intense flare at the beginning (segments II+III) to being steadier,
with weaker flare events (segments IV+V).  The light-curves also suggest
that the flare distribution is characterized by smaller values of $\alpha$
than are \fkaqr\ and \wolf.  Detailed analysis confirms the latter, with
$\overline{\alpha_{\rm II+III}}=2.17\pm0.03$ and
$\overline{\alpha_{\rm IV+V}}=2.31\pm0.11$
(see Figures~\ref{f:adleo_prob_23},\ref{f:adleo_prob_45})
but the distributions $p(\alpha)$ overlap for the two segments, and
we cannot rule out at the 10\% confidence level that the $\alpha$'s
are identical for the two datasets.  However, the trend for the
analyses of individual segments,
($\overline{\alpha_{\rm II}}=2.03\pm0.05$,
$\overline{\alpha_{\rm III}}=2.22\pm0.07$,
$\overline{\alpha_{\rm IV}}=2.21\pm0.06$, and
$\overline{\alpha_{\rm V}}=2.31\pm0.03$)
does suggest a gradual steepening of the flare distribution when
large flares are absent from the dataset.

%\clearpage

%\clearpage

%\clearpage

\section{Summary}
\label{s:summary}

We have modeled the event arrival times from active stars \fkaqr,
\wolf, and \adleo\ with particular attention to the component that
arises from flare-like events.  On the Sun, flares are known to be
distributed as a power\,law in energy (Hudson 1991; also see
Aschwanden et al.\ 2000 and references therein), and numerous studies
have established that strong stellar flares also follow a power\,law
distribution, with indices ranging from 1.5 - 2.5 (see Audard et
al.\ 2000 and references therein).  This is of considerable interest
because if the power\,law index $\alpha$ is $>2$, then the coronal
X-ray losses could in principle be ascribed to weak flare events
that are nevertheless numerous enough to dominate the emission.

We consider active dMe stars with known variability in their light
curves where numerous flares are seen.  Audard et al.\ (2000) find
that in general dF and dG stars tend to have $\alpha>2$ while dK
and dM stars tend to have lower $\alpha$.  In particular, they
analyze an older dataset of \adleo\ (from May 1996) and show that
detectable flares have $\alpha \in [1.18,2.35]$ or $\alpha=2.02\pm0.28$
using different methods.  A more detailed analysis by G\"{u}del et
al.\ (2002) based on a larger sample of the dataset used here shows
$2.0<\alpha<2.5$ .

We model the event arrival times using a simple two-component model
comprising of a constant rate component and a statistical ensemble
of flare components, with the flare energies distributed as a power\,law.
In general, the simplifying assumptions we make (e.g., constancy of
decay timescales, ignoring the rise times, assuming a constant
counts-to-energy conversion factor, including the background directly
in the model, etc.) are conservative, and tend to underestimate the
value of $\alpha$.  We find that all the stars in our sample clearly
have $\alpha>2$: for \adleo, $\alpha$ lies in the range 2.06 - 2.32;
for \fkaqr, $\alpha=2.60\pm0.34$; and for \wolf, $\alpha=2.74\pm0.35$.
We thus conclude that coronal heating on these stars is dominated by
impulsive energy release events whose energy output is
$\gtrsim~2-3~\times~10^{29}$ ergs, reaching to the microflare range.
These results are in contrast to the solar case, where over similar
flare energy ranges the observed distribution of flares is shallower,
with $\alpha \approx 1.8$, i.e., below the critical value of 2.

Further, we directly estimate the contribution of the flare emission
to the total observed count rate, as one of the parameters defining
the model.  Because the energy range over which the model is defined
spans over 4 orders of magnitude, and the power\,law indices indicate
steep distributions, we expect that the flare component should
contribute significantly to the total emission.  Indeed, we find this
to be generally $>50\%$, and in some cases being $>80\%$.  Thus, there
appears to be no truly ``quiescent'' emission on some of these low-mass
active stars, i.e., emission from apparently stable active region loops
as on the Sun is not a dominant component of the observed emission.

We have also explored the possible dependence of the various model
parameters on flare energies.  The long observation of
\adleo\ suggests that $\alpha$ increases when strong flares are
not evident in the data suggesting that the flare distribution
steepens for flares of smaller energies, though we cannot rule out
the possibility of a statistical fluctuation that mimics this trend.
We have also searched for, but do not find, evidence of strong
dependence of the decay timescale on flare energies.

We find that if the flare distributions extend to the microflare
regime, the energy output due to these weak flares is quite sufficient
to account for the entire coronal emission in the {\sl EUVE}/DS passband.
However, it must be noted that the error bars on the parameters
derived for the fainter stars \fkaqr\ and \wolf\ are quite large
(for instance, values of $\alpha<2$ cannot be completely ruled out
for \fkaqr, and a firm upper bound on $\alpha$ cannot be set for \wolf)
and it would be of considerable interest to verify and improve these
results (and also to extend them to a larger sample of stellar types)
using high-quality data such as those obtainable with {\sl Chandra}
and {\sl XMM-Newton}.
Data from these observatories are
characterized by good time resolution and in general larger count
rates, and will therefore allow us to explore the arrival time
difference distribution functions $g(\tdel)$ at smaller values of
$\tdel$, thereby extending the range of values of $\alpha$ that the
method is sensitive to.
%Such observations would also extend the analysis into the higher
%energy X-ray passband.

\acknowledgements

We would like to thank David van Dyk, Alanna Connors, Eric Kolaczyk,
and Olivia Johnson for useful discussions.  We also thank the referee,
J.\ Linsky, for valuable comments that improved the paper.
VLK was supported by NASA grants and the Chandra X-Ray Center during
the course of this research.  JJD was supported by the Chandra
X-Ray Center NASA contract NAS8-39073.
The PSI group acknowledges support from the Swiss National Science
Foundation (grants 2100-049343 and 2000-058827).
This research has made use of the SIMBAD database, operated at CDS,
Strasbourg, France.

%\clearpage

\appendix
\section{Glossary of terms \label{s:glossary}}

Here we compile, for reference, all the symbols used in
the text (Table~\ref{t:glossary}).

\begin{table*}[htb!]
\begin{center}
\caption{Glossary of terms \label{t:glossary}}
{\small
\begin{tabular}{l l l}
\tableline
\tableline
Symbol & Description & First Use \\
\tableline

$\Gamma(x)$ & the Gamma function, which for integer $x$ is $(x-1)!$ & \S\ref{s:algo}, Eqn.~\ref{e:likelihood} \\
$\Delta T$ & total duration of observation & \S\ref{s:model}, Eqn.~\ref{e:flrct} \\
$\Theta(\cdot)$ & Heaviside step function & \S\ref{s:model}, Eqn.~\ref{e:fcomp} \\
$\alpha$ & power-law index & \S\ref{s:intro}, Eqn.~\ref{e:powerlaw} \\
$\tdel$ & arrival time difference between two consecutive photons & \S\ref{s:algo} \\
$\kappa$ & normalization of the power-law in counts units & \S\ref{s:model}, Eqn.~\ref{e:cnorm} \\
$\phi(t)$ & Instrument correction factor that includes Primbsching, etc. & \S\ref{s:model}, Eqn.~\ref{e:model} \\
$\rho_i$ & the fraction of time a source spends at rate $r_i$ & \S\ref{s:algo}, Eqn.~\ref{e:dtdisum} \\
$\tau$ & flare decay time scale & \S\ref{s:model}, Eqn.~\ref{e:fcomp} \\
$\chi^2$ & $= \sum\limits_{i} \left(\frac{Data_i-Model_i}{error(Data_i)}\right)^2$, a statistic measuring the quality of a fit & \S\ref{s:algo}, Figure~\ref{f:dtdist} \\

%$A_{\rm{\mu}hs}$ & area (generally of active regions) in solar micro-hemispheres & \S\ref{s:assume} \\
$C(t)$ & model light curve, observable counts in $[t,t+dt]$ & \S\ref{s:model}, Eqn.~\ref{e:model} \\
$D$ & observed data & \S\ref{s:algo}, Eqn.~\ref{e:bayes} \\
$E$ & energy output of flare event & \S\ref{s:intro}, Eqn.~\ref{e:powerlaw} \\
$F_j$ & peak intensity of model flare $j$ & \S\ref{s:model}, Eqn.~\ref{e:fcomp} \\
${\bf M,M'}$ & set of parameters defining the model & \S\ref{s:model}, Eqn.~\ref{e:pars} \\
$N_f$ & total number of flares in the model & \S\ref{s:model}, Eqn.~\ref{e:fcomp} \\
$I$ & information necessary to define the problem & \S\ref{s:algo}, Eqn.~\ref{e:bayes} \\
$dN$ & number of flare events in $[E,E+dE]$ & \S\ref{s:intro}, Eqn.~\ref{e:powerlaw} \\

$c$ & counts due to a flare & \S\ref{s:model}, Eqn.~\ref{e:cnorm} \\
$c_{max}$ & maximum model counts possible due to a flare, for a given dataset & \S\ref{s:model}, Eqn.~\ref{e:dndc} \\
$c_{min}$ & minimum model counts due to a flare that is practicable to consider & \S\ref{s:model}, Eqn.~\ref{e:dndc} \\
$f(t)$ & flare model light curve intensity at time $t$ & \S\ref{s:model}, Eqn.~\ref{e:fcomp} \\
$g(\tdel)$ & frequency histogram of the distribution of $\tdel$ & \S\ref{s:algo}, Eqn.~\ref{e:dtdist} \\
$k$ & normalization of the power-law in energy units & \S\ref{s:model}, Eqn.~\ref{e:norm} \\
$p(A|B)$ & probability that statement $A$ is true given that statement $B$ is true & \S\ref{s:algo}, Eqn.~\ref{e:bayes} \\
$r(t)$ & model light curve intensity & \S\ref{s:model}, Eqn.~\ref{e:model} \\
$r_C$ & constant component model count rate & \S\ref{s:model}, Eqn.~\ref{e:par1} \\
$r_F$ & mean model count rate of flare component over duration of observation & \S\ref{s:model}, Eqn.~\ref{e:par1} \\
$r_T$ & mean total model count rate over duration of observation & \S\ref{s:model}, Eqn.~\ref{e:par2} \\

$var(x)$ & variance of quantity $x$ & \S\ref{s:algo}, Eqn.~\ref{e:varmean} \\
$\overline{x}$ & mean value of quantity $x$ & \S\ref{s:algo}, Eqn.~\ref{e:varmean} \\
$x_{_{MAP}}$ & maximum {\sl a posteriori} value of $x$, where $p(x)$ is maximum & \S\ref{s:algo}, Footnote~\ref{f:map} \\

\tableline
\end{tabular}
}
\end{center}
\end{table*}

%\clearpage

%\newpage


\begin{references}

\reference{} Ambruster, C.W., Sciortino, S., \& Golub, L.\ 1987, ApJS, 65, 273
\reference{} Aschwanden, M.J., \& Charbonneau, P.\ 2002, ApJ, 566, L000
\reference{} Aschwanden, M.J., \& Parnell, C.E.\ 2002, ApJ, 572, 1048
\reference{} Aschwanden, M.J., Tarbell, T.D., Nightingale, R.W., Schrijver, C.J, Title, A., Kankelborg, C.C., Martens, P., \& Warren, H.P.\ 2000, ApJ, 535, 1047
\reference{} Audard, M., G\"{u}del, M., \& Guinan, E.F.\ 1999, ApJ, 513, L53
\reference{} Audard, M., G\"{u}del, M., Drake, J.J., \& Kashyap, V.L.\ 2000, ApJ, 541, 396
\reference{} Benz, A.O., \& G\"{u}del, M.\ 1994, A\&A, 285, 621
\reference{} Butler, C.J., Rodono, M., Foing, B.H., \& Haisch, B.M.\ 1986, Nature, 321, 679
\reference{} Boffetta, G., Carbone, V., Giuliani, P., Veltri, P., \& Vulpiani, A.\ 1999, Phys.Rev.Lett., 83, 22, p4662
\reference{} Byrne, P.B., Butler, C.J., \& Lyons, M.A.\ 1990, A\&A, 236, 455
\reference{} Cheng, C.-C., Doschek, G.A., \& Feldman, U.\ 1979, ApJ, 227, 1037
\reference{} Collura, A., Pasquini, L., \& Schmitt, J.H.M.M.\ 1988, A\&A, 205, 197
\reference{} Crawford, D.F., Jauncey, D.L., \& Murdoch, H.S.\ 1970, ApJ, 162, 405
\reference{} Crosby, N.B., Aschwanden, M.J., \& Dennis, B.R.\ 1993, Sol.Phys., 143, 275
\reference{} Dempsey, R.C., Linsky, J.L., Schmitt, J.H.M.M., \& Fleming, T.A.\ 1993, ApJ, 413, 333
\reference{} Doyle, J.G., \& Butler, C.J.\ 1985, Nature, 313, 378
\reference{} Drake, J.J.\ 1996, {\sl Cool Stars, Stellar Systems, and the Sun}, 9$^{th}$ Cambridge Workshop ASP Conference Series, eds.\ R.Pallavicini \& A.K.Dupree, v109, p203
\reference{} Eadie, W.T., Drijard, D., James, F.E., Roos, M., \& Sadoulet, B.\ 1971, ``Statistical Methods in Experimental Physics'', North-Holland
\reference{} Georgoulis, M.K., \& Vlahos, L.\ 1998, A\&A, 336, 721
\reference{} Georgoulis, M.K., Vilmer, N., \& Crosby, N.B.\ 2001, A\&A, 367, 326
\reference{} Giampapa, M.S., Rosner, R., Kashyap, V., Fleming, T.A., Schmitt, J.H.M.M., \& Bookbinder, J.A.\ 1996, ApJ, 463, 707
\reference{} G\"{u}del, M. 1997, ApJ, 480, L121
\reference{} G\"{u}del, M., \& Benz, A.O.\ 1993, ApJ, 405, L63
\reference{} G\"{u}del, M., Guinan, E.F., \& Skinner, S.L.\ 1997, ApJ, 483, 947
\reference{} G\"{u}del, M., Audard, M., Guinan, E.F., Drake, J.J., Kashyap, V.L., Mewe, R., \& Alekseev, I.Y.\ 2001, in {\sl Proc. of X-Ray Astronomy 2000}, Palermo, Sep 2000, ASP Conf.Ser., Eds.\ R.Giacconi, L.Stella, and S.Serio, in press (astro-ph/0011572)
\reference{} G\"{u}del, M., Audard, M., Kashyap, V.L., Drake, J.J., \& Guinan, E.F.\ 2002, submitted to ApJ
%\reference{} Haisch, B.M., \& Schmitt, J.H.M.M.\ 1996, PASP, 108, 113
\reference{} Hudson, H.S.\ 1991, Sol.Phys., 133, 357
\reference{} Kashyap, V., \& Drake, J.J.\ 1998,  ApJ, 503, 450
\reference{} Kashyap, V., Rosner, R., Harnden, F.R.,Jr., Maggio, A., Micela, G., \& Sciortino, S.\ 1994, ApJ, 431, 402
\reference{} Krasnoselskikh, V., Podladchikova, O., Lefebvre, B., \& Vilmer, N.\ 2001, astro-ph/0104241, submitted to A\&A
\reference{} Katsukawa, Y., \& Tsuneta, S.\ 2001, ApJ, 557, 343
\reference{} Kellett, B.J., \& Tsikoudi, V.\ 1999, MNRAS, 308, 111
\reference{} Kopp, R.A., \& Poletto, G.\ 1993, ApJ, 418, 496
\reference{} Krucker, S., \& Benz, A.O.\ 1998, ApJ, 501, L213
\reference{} Kucera, T.A., Dennis, B.R., Schwartz, R.A., \& Shaw, D.\ 1997, ApJ, 475, 338
\reference{} Lepreti, F., Carbone, V., \& Veltri, P.\ 2001, ApJ, 555, L133
\reference{} Lin, R.P., Schwartz, R.A., Kane, S.R., Pelling, R.M., \& Hurley, K.C.\ 1984, ApJ, 283, 421
\reference{} Loredo, T.J.\ 1990, in ``Maximum Entropy and Bayesian Methods'', ed.\ P.F.\ Fougere (Dordrecht:Kluwer), 81
\reference{} Lu, E.T., \& Hamilton, R.J.\ 1991, ApJ, 380, L89
\reference{} Lu, E.T., Hamilton, R.J., McTiernan, J.M., \& Bromund, K.R.\ 1993, ApJ, 412, 841
\reference{} Moon, Y., Choe, G.S., Yun, H.S., \& Park, Y.D.\ 2001, at AGUSP, \#SP41A-01
\reference{} Narain, U., \& Ulmschneider, P.\ 1990, Sp.Sc.Rev, 54, 377
\reference{} Narain, U., \& Ulmschneider, P.\ 1996, Sp.Sc.Rev, 75, 453
\reference{} Norman, J.P., Charbonneau, P., McIntosh, S.W., \& Liu, H.-L.\ 2001, ApJ, 557, 891
\reference{} Osten, R., \& Brown, A.\ 1999, ApJ, 515, 746
\reference{} Pallavicini, R., Tagliaferri, G., Stella, L.\ 1990, A\&A, 228, 403
\reference{} Parker, E.N.\ 1988, ApJ, 330, 474
\reference{} Parnell, C.E., \& Jupp, P.E.\ 2000, ApJ, 529, 554
\reference{} Perryman, M.A.C., et al.\ 1997, A\&A, 323, 49
\reference{} Porter, J.G., Fontenla, J.M., \& Simnett, G.M.\ 1995, ApJ, 438, 472
\reference{} Reale, F., Peres, G., \& Orlando, S.\ 2001, ApJ, 557, 906
\reference{} Robinson, R.D., Carpenter, K.G., \ Percival, J.W., \& Bookbinder, J.A.\ 1995, ApJ, 451, 795
\reference{} Robinson, R.D., Carpenter, K.G., \& Percival, J.W.\ 1999, ApJ, 516, 916
\reference{} Robinson, R.D., Linsky, J.L., Woodgate, B.E., \& Timothy, J.G.\ 2001, ApJ, 554, 368
\reference{} Rosner, R., \& Vaiana, G.S.\ 1978, ApJ, 222, 1104
\reference{} Rosner, R., Golub, L., \& Vaiana, G.S.\ 1985, ARA\&A, 23, 413
\reference{} Schrijver, C.J., Title, A.M., Berger, T.E., Fletcher, L., Hurlburt, N.E., Nightingale, R.W., Shine, R.A., Tarbell, T.D., Wolfson, J., Golub, L., Bookbinder, J.A., Deluca, E.E., McMullen, R.A., Warren, H.P., Kankelborg, C.C., Handy, B.N., \& de Pontieu, B.\ 1999, Sol.Phys., 187, 261
\reference{} Shimizu, T. \& Tsuneta, S.\ 1997, ApJ, 486, 1045
\reference{} Shimizu, T.\ 1995, PASJ, 47, 251
\reference{} Skumanich, A.\ 1985, AuJPh, 38, 971
\reference{} Stepien, K., \& Ulmschneider, P.\ 1989, A\&A, 216, 139
\reference{} Strassmeier, K.G., Hall, D.S., Fekel, F.C., \& Scheck, M.\ 1993, A\&AS, 100, 173
\reference{} Temmer, M., Veronig, A., Hanslmeier, A., Otruba W., \& Messerotti, M.\ 2001, A\&A, 375, 1049 
\reference{} van Dyk, D., Connors, A., Kashyap, V.L., \& Siemiginowska, A.\ 2001, ApJ 548, 224
\reference{} Veronig, A., Temmer, M., Hanslmeier, A., Otruba, W., \& Messerotti, M.\ 2002, A\&A, 382, 1070
\reference{} Vlahos, L., Georgoulis, M., Kluiving, R., \& Paschos, P.\ 1995, A\&A, 299, 897
\reference{} Wheatland, M.S., Sturrock, P.A., \& McTiernan, J.M.\ 1998, ApJ, 509, 448 
\reference{} Wheatland, M.S.\ 2000, ApJ, 536, L109
\reference{} Winebarger, A., Emslie, A.G., Mariska, J.T., \& Warren, H.P.\ 2001, ApJ, in press

\end{references}
\end{document}